\documentclass[conference]{IEEEtran}

\usepackage{endnotes}

\usepackage{times}
\usepackage{cite}

\ifCLASSINFOpdf
\else
\fi

\usepackage[pdftex]{graphicx}

\usepackage{url}

\begin{document}


\title{Enhancing Application Performance by Memory Partitioning in Android Platforms}

\author{\IEEEauthorblockN{Geunsik Lim\IEEEauthorrefmark{1},
Changwoo Min\IEEEauthorrefmark{2} and
Young Ik Eom\IEEEauthorrefmark{3} } 
\IEEEauthorblockA{Sungkyunkwan University, Korea\IEEEauthorrefmark{1}\IEEEauthorrefmark{2}\IEEEauthorrefmark{3}         
                  Samsung Electronics, Korea\IEEEauthorrefmark{1}\IEEEauthorrefmark{2}}
\{leemgs\IEEEauthorrefmark{1}, multics69\IEEEauthorrefmark{2}, yieom\IEEEauthorrefmark{3}\}@skku.edu, \{geunsik.lim\IEEEauthorrefmark{1}, changwoo.min\IEEEauthorrefmark{2}\}@samsung.com
}



\maketitle

\begin{abstract}

This paper suggests a new memory partitioning scheme that can enhance process lifecycle, while avoiding Low Memory Killer and Out-of-Memory Killer operations on mobile devices. Our proposed scheme offers the complete concept of virtual memory nodes in operating systems of Android devices.


\end{abstract}
 



%
\IEEEpeerreviewmaketitle


\let\thefootnote\relax\footnote{This work was supported by the IT R\&D program of MKE/KEIT [10041244, SmartTV 2.0 Software Platform]. This research was supported by Basic Science Research Program through the National Research Foundation of Korea(NRF) funded by the Ministry of Education, Science and Technology [2011-0025971].}

\section{Introduction}

Recent mobile phone users can use not only the built-in applications that the manufacturers included into the mobile phone, but also the third-party applications obtained from various app-markets. In these systems, due to the memory consumption of the third-party applications, there are frequent situations that the available memory space is insufficient to run those applications efficiently. Especially, in low-end mobile devices that do not have sufficient memory capacity, memory shortage may occur more frequently.

 
In this paper, we introduce a new memory partitioning scheme to get enhanced application performance during \textit{process lifecycle} \cite{lifecycle}, while avoiding Low Memory Killer (LMK) and Out-of-Memory Killer (OOMK) operations on mobile devices. We propose a complete memory partitioning framework at the operating system level.

The rest of this paper is organized as follows. In Section II, several technical issues on \textit{process lifecycle} are described. The new memory partitioning scheme for improving \textit{process lifecycle} is suggested in Section III. Section IV shows the evaluation results of the proposed scheme. Finally, Section V concludes the paper.

\section{Memory Management in Android Platform}

The operating system generally supports page reclamation \cite{pagereclaim}, swap in/out \cite{ramzswap}, \textit{cgroups} \cite{cgroups}, and OOMK \cite{oomrewrite} to settle the memory shortage problem. 

The page reclamation mechanism is useful to obtain available memory in the system. However, the mechanism always finds victim processes heuristically among the processes in the memory.

The mobile device manufacturers do not use the swap in/out technology in their commercial products because of the throughput issues of their applications.

Although the \textit{cgroups} provides a mechanism for aggregating/partitioning the set of tasks into several hierarchical groups, this mechanism does not prevent memory fragmentations because of the logical memory partitioning with private LRU of structure page cgroup per page.

The OOMK attempts to recover memory shortage from the OOM condition by killing low-priority processes \cite{pagereclaim} \cite{oomrewrite} which will most likely be the first victim. But, the operation of the OOMK results in the performance damage of new applications because of the thrashing occurred due to the limited memory resource of the mobile devices.

Android platform supports \textit{process lifecycle} mechanism to classify the processes based on the importance of the processes so that new applications can get the needed memory properly even when it reaches the situation of memory shortage. It controls the memory usage of each application via user-space components (Activity manager, Dalvik) and kernel-space components (LMK, OOMK) \cite{lifecycle} \cite{oomrewrite} to secure available memory stably.



Even under the system with large memory, memory shortage can happen when high-capacity and high-performance user applications come to run. Therefore, it is very important to secure as large available memory as possible. In Section III, we will describe our approach to solve this memory shortage problem on mobile devices.

\section{Virtual nodes to avoid LMK operations}



\begin{figure}
\centering
\includegraphics[width=0.9 \columnwidth,height=1.5in]{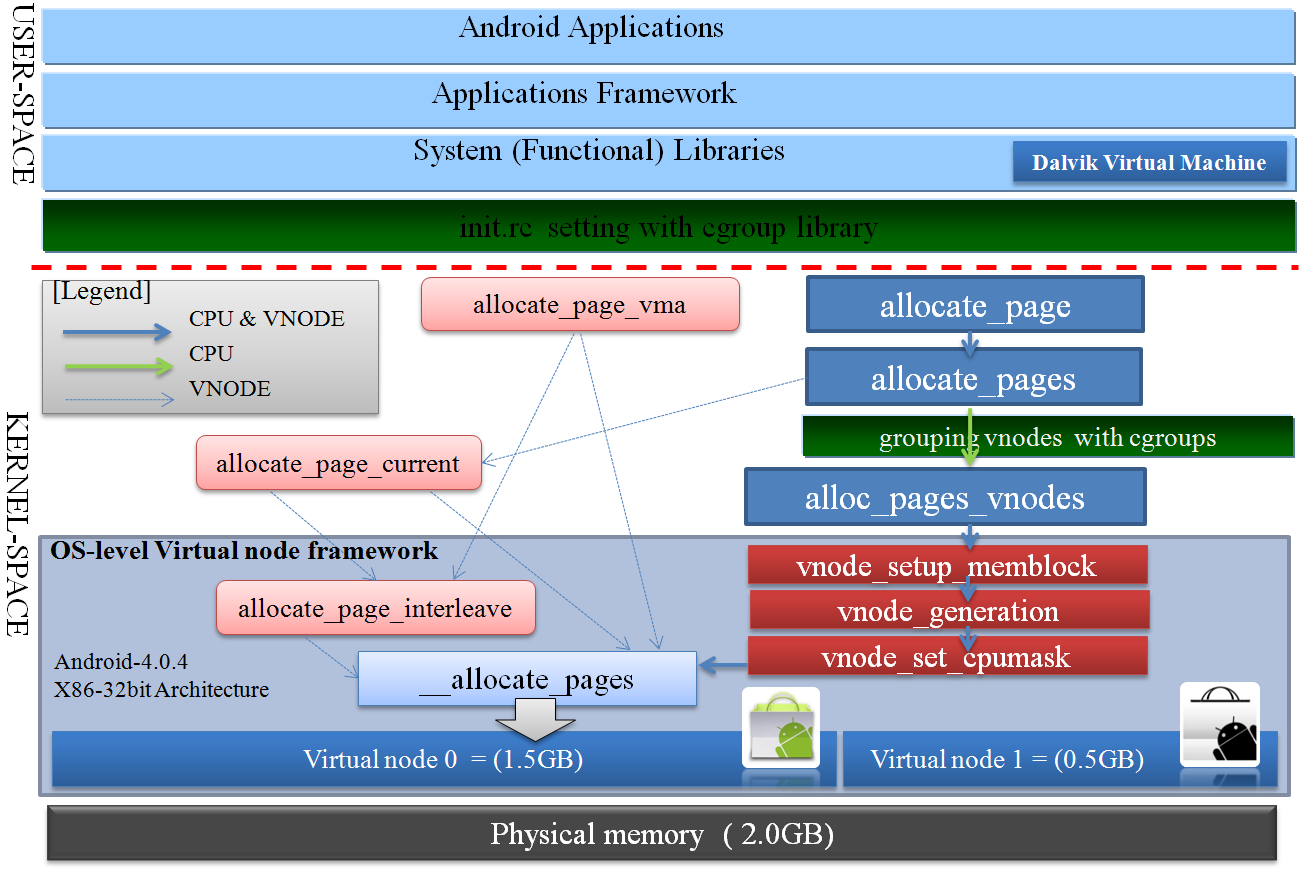}
\caption{The architecture of proposed memory partitioning}
\label{lim-fig-architecture}
\end{figure}

Figure \ref{lim-fig-architecture} shows the overall architecture of the new memory partitioning technique for improving \textit{process lifecycle} of the Android platform with the physically limited memory space. Our proposed memory partitioning technique mainly consists of three components as follows:
\begin{enumerate}
\item
\emph{vnode\_setup\_memblock}: sets up a memory node virtually from the start address to the end address.
\item
\emph{vnode\_generation}: generates a physical memory configuration that maps between the virtual node and the physical memory address, and determines the size of the physical distance table.
   \begin{enumerate}
   \item virtual node is a communication channel for logically partitioned memory access between the virtual memory and the physical memory.
   \item physical distance table is a map to get the physically separated memory block.
   \end{enumerate}
\item
\emph{vnode\_set\_cpumask}: allocates specific CPU masks for mapping between each CPU and each virtual node.
\end{enumerate}

Our design has two advantages in Android-based mobile devices: (1) limiting the memory consumption of untrustworthy applications by partitioning the memory space into two areas, virtual node (VNODE) 0 for reliable applications (official market) and virtual node (VNODE) 1 for unreliable applications (black market), (2) avoiding LMK and OOMK operations which happen under physical memory shortage.

The arrows of Figure \ref{lim-fig-architecture} represent the operations on the CPU and memory when an administrator sets the virtual memory nodes of the operating system from a physical memory on Android devices. For example, we run some critical applications only in VNODE 0. Also, we run other applications in VNODE 1. Through this approach, the operating system manages applications to avoid reaching the memory shortage even in a long running system.

Our memory partitioning scheme prevents an application from exhausting the entire memory by executing critical applications only in VNODE 0. Accordingly, these critical applications will stay in the memory of VNODE 0 continuously until a user terminates the critical application.

The key idea is that non-critical applications run in the physically partitioned specific memory area. This operation helps the system to avoid reaching no free memory. These non-critical applications only return their allocated memory with the page reclamation algorithm of Linux. 


The proposed system completely offers virtual memory nodes at the operating system level for enhanced \textit{process lifecycle} in Android devices. This equipment supports scalable system infrastructure as follows:
\begin{itemize}
\item
Virtually separated memory space.
\item
Operating system level memory isolation.
\item
Advanced page reclamation based on virtual nodes.
\item
Memory controller interface at boot time.
\end{itemize}

\section{RESULTS}

We ported the latest \textit{Android Ice Cream Sandwich 4.0.4} and \textit{Busybox 1.18} to \textit{Samsung SENS R60+ (CPU: Intel Core2Duo, MEM: DDR2 2G)} laptop to verify that our technical approach can be effective on the Android mobile platform. We also booted the \textit{Android Ice Cream Sandwich} including our new memory partitioning scheme based on Linux kernel 3.0 as a test bed for the Android tablet platform. We configured the system by creating two virtual memory nodes in different size, \textit{“VNODE 0 of 1.5 GB and VNODE 1 of 0.5 GB”}.   

\begin{figure}
\centering
\includegraphics[width=0.9 \columnwidth,height=1.4in]{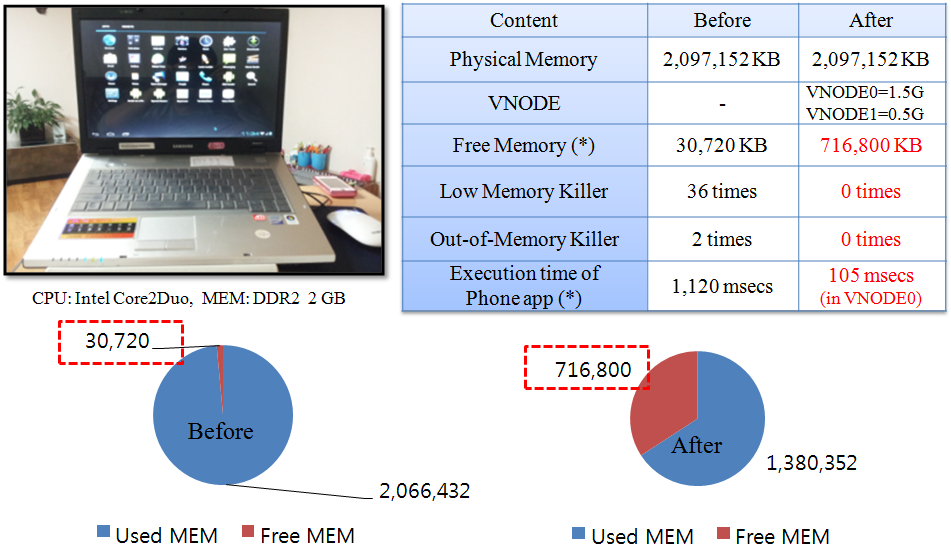}
\caption{The available memory result with virtual memory node}
\label{lim-fig-evaluation}
\end{figure}

We evaluated and compared the memory consumption of the existing approach (\textit{before}) and the proposed approach (\textit{after}) when we executed the sequential file I/O operation with the raw contents of the size of 1.5 GB into VNODE 1 for 2 days. Figure \ref{lim-fig-evaluation} shows the available memory size, the result of LMK, and the status of the OOMK after running the sequential file I/O operations. 

From our experiments, we gained the additional free memory of 670 MB and the reduced \textit{Phone} application execution time of 1,015 milliseconds over the existing systems. Since the test workload on VNODE 1 can use only 0.5 GB memory, Linux kernel executes many page reclamation operations in VNODE 1. Through our approach, the proposed system does not meet the operation of LMK and/or OOMK which operates on free memory shortage, 335 MB in our experimental environment. The frequencies of the execution of memory killers, both LMK and OOMK, were improved dramatically after adjusting the virtual nodes based on the new memory partitioning scheme.

\section{Conclusion}


We proposed a virtual memory node technique for memory partitioning. It focuses on the page reclamation operation of non-critical applications and the non-page reclamation operation of critical applications. Also, our approach supports virtual memory isolation to separately run applications of black markets and applications of official markets in the Android platform based on discontiguous memory access model. These approaches prevent LMK and OOMK from killing processes because of the memory shortage of the system. 

In conclusion, our approach innovatively overcomes the poor performance of applications incurred due to the operations of LMK and OOMK, without any physical memory extension.

\bibliographystyle{IEEEtran}
\bibliography{ref}

\end{document}